\documentclass[conference]{IEEEtran}
\IEEEoverridecommandlockouts
\usepackage{cite}
\usepackage{amsmath,amssymb,amsfonts}
\usepackage{algorithmic}
\usepackage{graphicx}
\usepackage{textcomp}
\usepackage{xcolor}
\usepackage{algorithmic}
\usepackage[ruled,vlined,linesnumberedhidden]{algorithm2e}
\def\BibTeX{{\rm B\kern-.05em{\sc i\kern-.025em b}\kern-.08em
    T\kern-.1667em\lower.7ex\hbox{E}\kern-.125emX}}
\begin{document}

\title{Low-Latency SCL  Bit-Flipping Decoding of Polar Codes }
\author{
	\IEEEauthorblockN{Wei Zhang and  Xiaofu Wu}
	\IEEEauthorblockA{Nanjing University of Posts and Telecommunications, Nanjing 210003, China}
	\IEEEauthorblockA{1021010419@njupt.edu.cn; xfuwu@ieee.org}
}

\maketitle

\begin{abstract}
 Bit flipping can be used as a postprocessing technique to further improve the performance for successive cancellation list (SCL) decoding of polar codes. However, the number of bit-flipping trials could increase the decoding latency significantly, which is not welcome in practice. In this paper, we propose a low latency SCL bit flipping decoding scheme, which is restricted to just single round of post-processing. The use  of multiple votes for a more accurate estimation of path survival probability is proposed to locate the first error event of SCL decoding. Simulations show the sound improvement compared to the existing SCL bit-flipping decoding methods. 
\end{abstract}


\section{Introduction}
Polar codes in combination with successive cancellation (SC) decoding have been theoretically demonstrated to have capacity-achieving capabilities \cite{Ref_1} under the binary-input discrete memoryless channel (BI-DMC). However, the performance of SC decoding is unsatisfactory for practical short codes. In order to overcome this deficiency of SC decoding, successive cancellation list (SCL) \cite{Ref_2} decoding was proposed that preserves multiple candidate paths for decision so as to increase the probability of successful decoding. Its performance  could be  further improved by using  Cyclic-Redundancy-Check-aided SCL (CA-SCL) \cite{Ref_3} decoding.

In recent years, bit-flipping was applied to either SC or SCL decoding for further improving the decoding performance. In \cite{Ref_4}, the so-called SC Flip decoding was first proposed, where the locations of the potential error-prone information bit positions are identified and a serial of re-decoding attempts are activated whenever its first attempt of decoding fails. In \cite{Ref_5}, the error-prone information bit positions were identified under the so-called critical set. \cite{Ref_6} proposed to flip multiple error-prone bits, which may improve the efficiency of bit-flipping.  

For SCL decoding, a number of bit-flipping strategies  \cite{Ref_7} \cite{Ref_8} following the concept of critical set were proposed for improving the performance of CA-SCL decoding.  In \cite{Ref_9},  a simple reliability metric was formulated with a full list of  path metrics  for locating the error prone bits, which shows its power for the purpose of bit-flipping, compared to \cite{Ref_7} \cite{Ref_8}.

Although varied SCL-Flip decoding schemes have achieved significantly improved performance than the standard SCL decoding. The number of re-decoding trials is rather large, which results in high-latency for the underlying decoder.  In this paper, we focus on the low-latency SCL-Flip decoding. By restricting a single re-decoding attempt, it is critical to improve the accuracy of locating the first error  position to be flipped. The idea of multiple votes for  locating the first error position is thus proposed. 

This paper is organized as follows, the preliminaries of polar codes are briefly introduced. Low-latency SCL bit-flipping decoding is described in detail in section III. Simulation results are given  in section IV. In section V, conclusion is made.

\section{PRELIMINARIES}
\subsection{Polar Codes}

Consider a $(N, K, \mathcal{A})$  polar code with block-length $N=2^n$, information bit length $K$ and information bit index set of $\mathcal{A}$.  Let  $u_1^N=(u_1,u_2,...,u_N)$  denote the input vector to be encoded, where $u_i$  is an information bit whenever  $i\in \mathcal{A}$ and a frozen bit for  $i\in \mathcal{A}^c$.  For polar encoding, $x_1^N=u_1^N\times G_N$ is employed with $G_N=F^{\otimes n}$, where $F^{\otimes n}$ denotes the $n$-th Kronecker power of $F=\left[ {\begin{array}{*{20}{c}}
		1&1\\
		1&0
\end{array}} \right]$.

\subsection{SC and SCL Decoding}
For SC decoding, it  successively evaluates the log-likelihood ratio of each bit $u_i$ based on the received vector $y_1^N$ and its $i$ preceding decision bits $ \hat{u}_{1}^{i-1}$
\begin{eqnarray}
	\mathrm{L}^i = \log \frac{P\left(y_1^{N},\hat u_1^{i - 1}|{u_i} = 0\right)}{P\left(y_1^{N},\hat u_1^{i - 1}|{u_i} = 1\right)}.
\end{eqnarray}
Then,  $\hat{u}_i$ is decided as 0 if $\mathrm{L}^i\geq 0$ and as 1 if $\mathrm{L}^i\leq 0$. 

Instead of just keeping  a single path, SCL preserves $L > 1$ paths during the decoding process, which could significantly improve the decoding performance. Let $\mathrm{L}_l^i$ denote the log-likelihood ratio of the bit $u_i$ along the $l$-th path. When the number of paths is greater than the list size as the SCL decoder proceeds from  level $i$ to $i+1$, it retains $L$ best paths according to the updated  path metric
\begin{flalign}
	PM_l^{i} =\left\{
	\begin{aligned}
		&PM_l^{i - 1},\quad \quad \quad \: {\text{if} \: {\hat u}_{i,l}} = \frac{1}{2}[1 - \text{sign}(\mathrm{L}_l^i)]\\ 
		&PM_l^{i - 1} + |\mathrm{L}_l^i|,\: \text{otherwise}.
		\\
	\end{aligned}
	\right.
\end{flalign}
where  $\text{sign}(x )=1$ if $x>0$ and $-1$ otherwise. 

%After all nodes  in  $\mathcal{A}$ are visited, the path with the smallest path metric is selected as the survival path.  For CA-SCL decoding,  the output $L$-paths are checked by CRC. Once a path passes the CRC check, it is claimed as the decoded output. Otherwise, a decoding failure is claimed.
After all nodes in $\mathcal{A}$ are visited, the paths in the list are examined one-by-one with decreasing metrics. The decoder outputs the first path passing the CRC detection as the estimation sequence. If none of such a path is found, the algorithm declares a decoding failure.

\subsection{SC-Flip and SCL-Flip Decoding}
SC-Flip decoding algorithm attempts to flip a decision to get
the correct decoding whenever the conventional SC decoding fails \cite{Ref_10}.
It was observed that error propagation occurs frequently in SC decoding, where any single erroneous
decision may result in a burst of errors. Hence,  it is crucial to find the first error position  in SC-Flip decoding.

For SCL decoding,  the bit-flipping can be again employed whenever a decoding failure is claimed.  The so-called SCL-Flip decoding was recently proposed in \cite{Ref_7, Ref_8, Ref_9}. In \cite{Ref_7}, a critical set is detected, which is deemed to be error-prone during SCL decoding. Therefore, each bit in this critical set is flipped in the re-decoding attempts. In \cite{Ref_9}, the bit position for flipping is determined by a newly-introduced confidence metric for the survival paths and simulations show significant performance improvement over the scheme in \cite{Ref_7}.

\section{Low-Latency SCL Bit-Flipping Decoding}
In \cite{Ref_11}, it was shown that the decoding failures in the CA-SCL decoding are mainly caused by the elimination of the correct path from $L$ maintaining paths.  In general, bit-flipping, as a post-processing technique, could be repeatedly implemented  if the previous attempt fails. However, the decoding latency increases linearly with the numbers of decoding attempts. Therefore,  this paper focuses on just single attempt of bit-flipping for the purpose of maintaining the  low-latency of SCL bit-flipping decoding.

%算法1
\begin{algorithm}[h]
	\caption{ Generation of  bit-flipping index set  $\mathcal{F}$}
	\textbf{procedure} $GenFlip\left(\{E_i(\alpha_1)\},\{ E_i(\alpha_2)\}\right)$ \\	
	\KwIn{$\{E_i{(\alpha_1)}, i\in \mathcal{A}\setminus \mathcal{A}_0\}$, $\{E_i{(\alpha_2)}, i \in \mathcal{A}\setminus \mathcal{A}_0\}$ }
	%\quad \quad $\{E_i{(\alpha_1)}, i\in \mathcal{A}\setminus \mathcal{A}_0\}$,  $\{E_i{(\alpha_2)}, i \in \mathcal{A}\setminus \mathcal{A}_0\}$\\
	\KwOut{$\mathcal{F}$}
	%\quad \quad $\{E_i{(\alpha_1)}, i\in \mathcal{A}\setminus \mathcal{A}_0\}$,  $\{E_i{(\alpha_2)}, i \in \mathcal{A}\setminus \mathcal{A}_0\}$\\
	
	$\mathcal{E}^{\alpha_1}\leftarrow E_i{(\alpha_1)}$, $\mathcal{E}^{\alpha_2}\leftarrow E_i{(\alpha_2)}$\\
	$\mathcal{I}_1 \leftarrow sort(\mathcal{E}^{\alpha_1})$, $\mathcal{I}_2 \leftarrow sort(\mathcal{E}^{\alpha_2})$\\
	\For{$m=1$ to $Q$}{
		$\mathcal{F}\leftarrow \mathcal{I}_1(1:m)\cap \mathcal{I}_2(1:m)$\\
		\If{$\mathcal{F} \neq \emptyset $}{
			break;
		}
	} 
	return $\mathcal{F}$\\
	\textbf{end procedure}
\end{algorithm}

\begin{algorithm}[h]
	\caption{Low-latency SCL-Flipping decoding}
	\KwIn{the received LLRs $y_1^N$, code length $N$, list size $L$, non-frozen set $\mathcal{A}$}
	\KwOut{the recovered message bits  $\hat{u}_1^N$}
	\SetKwFunction{Fsubroutine }{subroutine $SCLF(y_1^N,\, \mathcal{A}^c\, ,L\, ,\, pre)$}
	$i_1 \leftarrow -1$  $//$predicted position\\ 
	$crcPASS\leftarrow false$\\
	$\left(\hat{u}_1^N[1..L], \{E_i(\alpha_1)\},\{ E_i(\alpha_2)\}\right)$$\leftarrow SCLF(y_1^N, L, i_1)$\\
	\For{$l\leftarrow 1\: to\: L$}{
		\If{$CRC(\hat{u}_1^N[l])=success$}{
			$\hat{u}_d\leftarrow \hat{u}_1^N[l]$\\
			$crcPass\leftarrow true$
		}
	}
	\If{$crcPASS=false$}{
		%	\For{$i \leftarrow 1\: to\: 1$}{
			$\mathcal{F} \leftarrow GenFlip\left(\{E_i(\alpha_1)\},\{ E_i(\alpha_2)\}\right)$ \\	
			$i_1\leftarrow$ the first member of $\mathcal{F}$\\
			$\hat{u}_1^N[1..L]\leftarrow SCLF(y_1^N,L,i_1)$\\
			\For{$l\leftarrow 1\: to\: L$}{
				\If{$CRC(\hat{u}_1^N)=success$}{
					$\hat{u}_d\leftarrow \hat{u}_1^N[l]$\\
					$crcPass\leftarrow true$\\
					break
				}
			}
			\If{$crcPASS=true$}{
				break
			}
			%	}
	}
	\Return {$\hat{u}_d$}\\
	% Set Function Names
	\SetKwFunction{Fsubroutine}{subroutine $SCLF(y_1^N, L, i_1)$}
	\SetKwProg{Fn}{}{:}{\KwRet}
	\Fn{\Fsubroutine}{
		\For{$n\leftarrow 1\, to\, N$}{
			Perform one standard SCL Decoding:\\
			Path pruning when $n\in \mathcal{A} $ AND $L<2^n$\\
			\If{$n\neq index$}{
				$List\leftarrow \{1,...,L\}$\\
			}
			\Else{
				$List\leftarrow \{L+1,...,2L\}$\\
				%	update $1 \: to \: n-1$ information bits corresponding to the last L paths\\	
				%	update $1 \: to \: n-1$ $PM$ corresponding to the last L paths					
			}
		}
		\Return $\hat{u}_1^N[1..L]$
	}
\end{algorithm}

\subsection{Identification of the First Error Position with Multiple Votes}
In \cite{Ref_9}, it was shown that the confidence in the decision for the path competition on $u_i,  i \in \mathcal{A}\setminus {\mathcal{A}_0}$, can be determined
from the ratio between the total probability of the $L$ survival paths to the total probability of the $L$ removed paths, namely, 
\begin{eqnarray}
	\label{eq:conf}
	{E_i}(\alpha ) = \log \frac{{\sum\nolimits_{l = 1}^L {{e^{ - PM_l^{i}}}} }}{{{{\left( {\sum\nolimits_{l = 1}^L {{e^{ - PM_{l + L}^{i}}}} } \right)}^\alpha }}}
\end{eqnarray}
in the case of $\alpha\ge 1$. Note that $\mathcal{A}_0$ denotes the set consisting of $\log_2 L$ information indices.  However, once the first decoding failure at level $i_1 \in \mathcal{A}$ occurs during  CA-SCL decoding, it is likely to produce  error propagation in the subsequent decoding process for $i > i_1$.   In this case, there will be a biased estimate in the confidence of the decision which is less than its correct value for a large index $i$. To compensate for the biased estimate due to the error propagation, $\alpha\ge 1$ is introduced in (\ref{eq:conf}).  

Essentially, an information index $i$ that has a low $E_i(\alpha)$ should have a high priority for re-decoding. Therefore, a bit-flipping index set is constructed in \cite{Ref_9} by locating the bit indices with the smallest values of $E_i(\alpha)$. 

Since we are interested in the first error position, this may simply adopt the following rule, namely, 
\begin{eqnarray}
	\label{eq:min}
	\hat{i}_1 = \min_{i} E_i(\alpha),   i\in \mathcal{A}\setminus \mathcal{A}_0. 
\end{eqnarray}

However, whenever  a decoding failure occurs,  there may exist multiple bit errors in the final decoded output $\hat{u}_1^N$.  This means that  $Q> 1$ error positions $\{i_1,\cdots,i_Q\}\subset \mathcal{A}$ may appear in $\hat{u}_1^N$.  All of these error positions may have low confidence in $E_{i_q}(\alpha), q\in [1,Q]$. This makes the location of the first error position using (\ref{eq:min}) unreliable. 

Therefore, we employ the idea of multiple votes for better locating the first error position of CA-SCL decoding.  Consider the CA-SCL decoding with list size of $L$. For $i\in \mathcal{A}\setminus \mathcal{A}_0$, each decoding path $\hat{u}_{1,l}^{i-1}$, $\forall l\in[1,L]$, will extend to two paths where $\hat{u}_i=$0 or 1.

Firstly, we employ $\alpha_1\ge 1$ for evaluating the confidence of deciding each position $i \in \mathcal{A}$. If the first error position occurs at  $i_1 \in \mathcal{A}$, it is more likely that $E_{i}(\alpha_1) \ge E_{i_1}(\alpha_1), \forall i <i_1,  i \in \mathcal{A}$.  Due to the possible error propagation, this may be violated whenever the CA-SCL decoder proceed to the later position at $i> i_1$. Hence, we require an additional vote for eliminating the possible location with low value of $E_i(\alpha_1)$ but with $i>i_1$. This is pursued with the use of  $E_i(\alpha_2)$, $i\in \mathcal{A}$ with a large value of $\alpha_2\neq \alpha_1$. With the use of $\alpha_2 > 1$, the error position of $i>i_1$ may be more accurate since the bias of the deciding confidence ($E_i(\alpha_1)$) can be well alleviated.  With two votes for deciding the first error position, we believe that the estimation accuracy should be improved. In what follows, we provide a detailed algorithm for locating the first error position by employing two weighting factors, $\alpha_1$ and $\alpha_2$ in formulating the deciding confidences.  

Define two confidence sets as follows
\begin{eqnarray}
	\left\{
	\begin{aligned}
		\mathcal{E}^{\alpha_1}=\{E_i(\alpha_1)\,|\, i\in \mathcal{A}\setminus \mathcal{A}_0\}, \\
		\mathcal{E}^{\alpha_2}=\{E_i(\alpha_2)\,|\, i\in \mathcal{A}\setminus \mathcal{A}_0\}.
	\end{aligned}
	\right.
\end{eqnarray}

By sorting the above two sets independently, the index of elements in sorted ascending order  can be obtained as
\begin{eqnarray}
	\left\{
	\begin{aligned}
		\mathcal{I}_1=sort(\mathcal{E}^{\alpha_1})\\
		\mathcal{I}_2=sort(\mathcal{E}^{\alpha_2})
	\end{aligned}
	\right.
\end{eqnarray}
where $sort(\cdot)$ returns the indices of $ \mathcal{E}^{\alpha}$ with the ascending order of  $E_{i_1}^{\alpha}\leq E_{i_2}^{\alpha}\leq...\leq E_{i_Q}^{\alpha}$.

Then, we can locate the first error position by searching over 
\begin{eqnarray}
	\mathcal{F}_m =\mathcal{I}_1(1:m)\cap \mathcal{I}_2(1:m)
\end{eqnarray}
from $m=1$ to $Q$. Whenever $\mathcal{F}_m \neq \emptyset$, it succeeds. The corresponding algorithm is summarized as Algorithm 1. Note that we employ a set of $\mathcal{F}$ for the output, which facilitates the generalization to the case of $|\mathcal{F}|\ge 1$. 

\begin{figure}[ht]
	\centering
	\includegraphics[width=80mm]{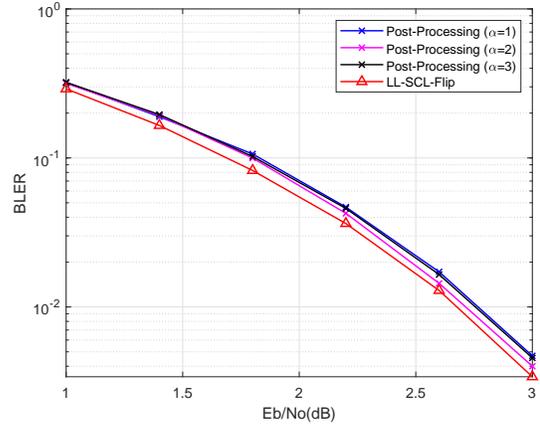}
	\caption{BLER performance comparison between LL-SCL-Flip and Post-Processing [7] with  different $\alpha$'s and list size of  $L=4$.}
	\label{fig:ListSize}
\end{figure}

\begin{figure}[ht]
	\centering
	\includegraphics[width=80mm]{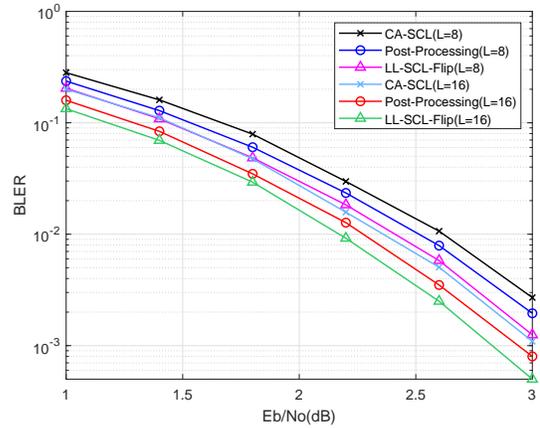}
	\caption{BLER performance  comparison among six  kinds of SCL decoding algorithms for PC(128,64+8). The list size is set as $L=8$ and $L=16$.}
	\label{fig:ListSize}
\end{figure}

\begin{figure}[ht]
	\centering
	\includegraphics[width=80mm]{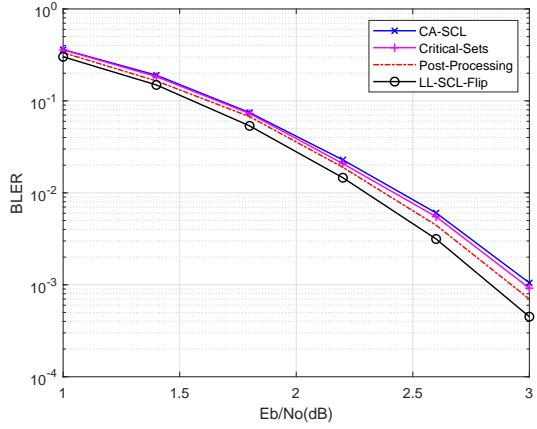}
	\caption{BLER performance comparison among four kinds of SCL decoding algorithms for PC(256,128+8). The list size is set as $L=4$.}
	\label{fig:ListSize}
\end{figure}

\begin{figure}[ht]
	\centering
	\includegraphics[width=80mm]{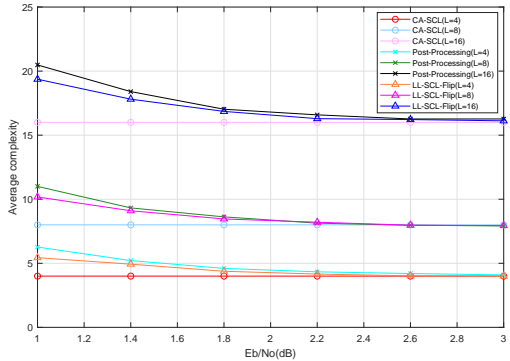}
	\caption{Average complexity of decoding for PC(128,64+8).}
	\label{fig:ListSize}
\end{figure}

\subsection{Low Latency SCL-Flip Decoding}
As a post-processing technique for SCL decoding,  the proposed low-latency SCL-Flip  (LL-SCL-Flip) takes just one chance of bit-flipping after the standard SCL decoding.  Whenever SCL decoding fails, it initiates a new re-decoding attempt, where SCL decoder restarts its decoding from the error position $i_1$ with the shift-pruning approach \cite{Ref_9}.  Instead of using $L$ best paths, the re-decoding attempt tries to employ the discarded $L$ paths, since the confidence of deciding $u_{i_1}$ is low according to multiple-votes investigated in Algorithm 1. The proposed LL-SCL-Flip is summarized in Algorithm 2.

\section{Simulations}
In this section,  simulations are performed to show the effectiveness of the proposed LL-SCL-Flip decoding, which is compared with various existing SCL bit-flipping decoding, including the partial SCL-Flip \cite{Ref_7} and Post-Processing \cite{Ref_9}. A (128, 64+8) polar code with 8-bit CRC is considered, where its  generator polynomial is  $x^8+x^7+x^6+x^5+x^4+x^3+x^2$. All the code bits are BPSK $(1\to-1, 0\to1)$ modulated and transmitted over an AWGN channel ($\sigma = \frac{1}{\sqrt{2R}}10^{-\frac{snr}{20}}$). 

\subsection{BLER Performance}
First, we analyze the decoding performance.Fig. 1 shows the performance of  the proposed LL-SCL-Flip decoding ($L=4$) for polar codes (128, 64+8), which is compared with Post-Processing with $\alpha=1$, $\alpha=2$, $\alpha=3$ in the case of just one-chance of re-decoding. 

For the polar code of (128, 64+8), Fig. 2 shows  that  LL-SCL-Flip with $L=8$ outperforms  the standard CA-SCL($L=8$) by about 0.25 dB and is inferior to the standard CA-SCL($L=16$) by about 0.02 dB. Compared to the just one-round of post-processing ($\alpha=2$), it achieves about 0.12 dB gain. For a larger list size  $L=16$, Fig. 2 shows  that LL-SCL-Flip outperforms the standard CA-SCL ($L=16$) by about 0.21 dB. Compared to Post Processing with one single round ($\alpha=2$), it achieves about 0.11 dB gain.

For the polar code of (256, 128+8), Fig. 3 shows that the proposed LL-SCL-Flip with $L=4$ is also very effective, which gains the improvement about 0.19 dB  over that of the standard CA-SCL, and 0.1 dB over the one-round of post-processing ($\alpha=2$) in \cite{Ref_9}. The performance of LL-SCL-Flip is better than the partial SCL-Flip \cite{Ref_7}.

\subsection{Average Complexity of Decoding}

In this paper, we use the average list size to represent the decoding complexity \cite{Ref_6}, and let $F$ represent the total number of decoding frames and $T$ represent times. For example, the average complexity of the CA-SCL decoding algorithm is $(L\cdot T/F)$. For the bit-flipping decoding algorithm, each additional flip means that the complexity is increased by 1. If the proposed flip decoding algorithm tries to flip additional $t$ times, the complexity is $(T+t)\cdot L/F$.

In Fig. 4, we compare the average decoding complexity of the standard CA-SCL, Post-Processing \cite{Ref_9} and the proposed LL-SCL-Flip. Fig. 4 shows that the average computational complexity of LL-SCL-Flip is a bit higher than either Post-Processing \cite{Ref_9}, or standard CA-SCL decoding before Eb/No=2.4dB. They are almost alike after Eb/No=2.4dB.
\section{Conclusion}
This paper considers to reduce the latency of the standard bit-flipping approach for SCL decoding. Instead of using a large number of re-decoding attempts, we focus on the single-round approach for bit-flipping. By employing the idea of multiple votes, this paper proposed an enhanced mechanism for locating the first error position, with which a shift-pruning approach is used in the re-decoding. Simulations show that the proposed LL-SCL-Flip could achieve obvious improvement over the standard SCL-Flip decoding.

\end{document}